\documentclass[12pt]{article}
\usepackage{amsmath}
\usepackage{graphicx}
\usepackage{color}
\usepackage[margin = 1 in]{geometry}
\begin{document}
	{\color{black}
\begin{center}
{\large \bf Investigation of the Dunkl-Schr\"odinger equation for Position Dependent Mass in the presence of a Lie algebraic approach}
\end{center}
}
\vspace{.5cm}
\begin{center}
	{\bf P. Sedaghatnia$^{\dagger}$},\,\,\,
	{\bf H.  Hassanabadi $^{\dagger,\ddagger, }$}\footnote{h.hasanabadi@shahroodut.ac.ir (Corresponding author)}
		{\bf W. S. Chung$^{\S}$},\,\,\,
	{\bf B. C. L\"utf\"uo\u{g}lu $^{\ddag,\$}$},\,\,{\bf S. Hassanabadi $^{\ddag}$} and
	{\bf J. K\u{r}\'\i\u{z}$^{\ddag}$}\,\,\,    \\
	\small \textit {\it $^{\dag}$ Faculty of Physics, Shahrood University of Technology, Shahrood, Iran \\ P. O. Box : 3619995161-316.}\\
\small \textit {\it $^{\ddag}$ Department of Physics, University of Hradec Kr$\acute{a}$lov$\acute{e}$,
		Rokitansk$\acute{e}$ho 62, 500 03 Hradec Kr$\acute{a}$lov$\acute{e}$, Czechia.}\\	
\small \textit {\it $^{\S}$ Department of Physics and Research Institute of Natural Science,\\
		College of Natural Science,\\
		Gyeongsang National University, Jinju 660-701, Korea}\\
	\small \textit {\it $^{\$}$ Department of Physics, Akdeniz University, Campus 07058, Antalya, Turkey.}\\

\end{center}
{\color{black}
\begin{abstract}
Recent studies have shown that the use of Dunkl derivatives instead of ordinary derivatives leads to deriving parity-dependent dynamic solutions. According to this motivation in this manuscript, we formulate the Dunkl-Schr\"odinger equation within the position-dependent mass formalism and derive an algebraic solution out of it.  Our systematic approach lets us observe some new findings in addition to the earlier ones. For example, we find that the solution of the Dunkl-Schr\"odinger equation with position-dependent mass cannot be considered independent from the choice of parameters. Similarly, through the sl(2) algebra, the energy spectrum and the corresponding wave functions are derived in terms of possible Dunkl, ($\mu$), and mass,  ($\alpha$), parameters.
\end{abstract}
}
\section{Introduction}
{\color{black}
Quantum mechanics and its algebra receive much attention in physics since the very beginning of the last century. In this study, we take the Dunkl derivative as an operator of  quantum mechanics \cite{D1,D2,D3,D4,D5,D6,D7,D8,D9,D10,D11,D12,D13}.
The considered Dunkl derivative consists of three parts. One of them has the normal derivative term, while the other parts include parity, which can be written in two states, namely even parity, and odd parity states. In fact, the Dunkl operator is sometimes called the general form of the initial theory since it possesses modifications of the ordinary derivative. Therefore, in the limit case of the Dunkl deformation parameter, the initial form of the normal derivative has to be recovered. One of the most interesting and important properties of the Dunkl operator arises when it is used in the Schrödinger equation. To be more precise, it leads to a square root term that is not in the ordinary case, and for a special value of $n$, this square root term vanishes the degeneracy. This means that in the ordinary case, for a special quantum number, the state of the system can only be odd or even. But in the presence of the square root term, we have both odd and even case parities as the separate cases. Also, the construction of algebra, which includes the operators of creation and annihilation that are proportional to the degrees of freedom, will lead to the generalization of field theory in this field. }

On the other hand, position-dependent mass formalism, hereafter PDM, is a broad approach applicable to a wide range of scientific fields and has been the subject of many recent studies \cite{PDM1,PDM2,PDM3,PDM4,PDM5,PDM6}. For example, non-relativistic particle dynamics has been investigated previously by solving the Schr\"odinger equation with the PDM formalism under various approaches such as path integral \cite{PDM7}, supersymmetric quantum mechanics \cite{PDM8}, and Hamiltonian factorization methods \cite{PDM9}. In a recent work, authors suggested that Shannon and Fisher  information entropy  could be used  alternatively to ordinary entropy \cite{Can1}. The attempt of linking the PDM formalism to the alternative entropy is also noticeable \cite{PDM10, Can2}. For example, in \cite{PDM10} authors discussed the Fisher information of a particle with a nonuniform solitonic mass density after they solved the PDM Schr\"odinger equation with a squared hyperbolic co-secant potential energy. In an interesting paper related to the classical oscillator quantization, the authors in \cite{PDM11} thought of operator equivalents of momentum and PDM in the same symmetrical way and obtained a quantum Hamiltonian that is Hermitian in type. Very recently, author discussed the strong gravitational effects on the energy of the particle trapped in an infinite square well with the PDM formalism \cite{Can3}.

According to these facts, we believe that examining the PDM formalism in the Dunkl-Schrödinger equation, which considers parity as well, is going to give a great contribution to the field. Up to our best knowledge, no study along this line has been carried on so far. With this motivation we prepare the rest of the paper as follows: In section \ref{sec2}, we briefly introduce Quasi-exactly solvable (QES) differential equations within general terms. Then, in Section \ref{sec3} We construct the system that contains a hidden sl(2) algebraic structure  which is responsible for the QES structure. We obtain exact expressions for the energies and corresponding wavefunctions as well as the allowed values for the potential parameters from the representation theory of sl(2).  In Section \ref{sec4}, we finish the work with our conclusion.

\section{Quasi-exactly solvable differential equations}\label{sec2}

In a Hilbert space, $H$ is a differential linear operator of the QES equation  \cite{QES0,QES1,QES2,QES3,QES4}
\begin{equation}
(H - E)\Psi  = 0,\quad\quad 
{H_0} = H - E, \quad \quad
{H_{_0}}\Psi  = 0,
\end{equation}
where $\tilde {H} $  equates to $ H_{0} $ under the transformation of
\begin{equation}\label{2.2}
\tilde H = {G^{ - 1}}.H_{0}.G 
\end{equation}
It should be noted that the transformed operator $\tilde{H}$ keeps the eigenvalue of $\tilde{H}\tilde{\psi}=E \tilde{\psi}$. For the first-order differential operators in one dimension, the only Lie algebra with a finite-dimensional representation is the sl(2) algebra, whose generators are in the form of:
%\begin{equation}\tag{A.7\color{black}}
%({P_4}\frac{{{d^2}}}{{d{r^2}}} + {P_3}\frac{d}{{dr}} + {P_2})\tilde \Psi (r) = 0
%\end{equation}
\begin{equation}\label{2.3}
J_n^ +  = {x^2}\frac{d}{{dx}} - nx, \quad\quad
J_n^0 = x\frac{d}{{dx}} - \frac{n}{2}, \quad\quad
J_n^ -  = \frac{d}{{dx}},
\end{equation}
where $ \tilde H $ can be expressed as a quadratic combination of sl(2) generators
\begin{equation}\label{2.4}
\tilde H = \sum\limits_{a,b = 0, \pm } {{C_{ab}}{J^a}{J^b} + \sum\limits_{a = 0, \pm } {{C_a}{J^a} + C} } 
\end{equation}
The latter operator is not a Schrodinger-like operator, but by changing the variable and going through a gauge transformation it always can be reduced to a Schrodinger-type operator. In the next section we are going to employ this method and obtain the eigenvalues and the corresponding eigenfunctions.

\section{Algebraic solutions of the Dunkl-Schrodinger equation with PDM}\label{sec3}
Let us begin the analysis by expressing the position-dependent mass Schrodinger equation \cite{1p, 2p} in the presence of the Dunkl derivative 
\begin{eqnarray}\label{1}
\bigg[-\frac{\hbar^{2}}{2}D_{x}^{\mu }\bigg(\frac{1}{\sqrt{m(x)}}D_{x}^{\mu }\frac{1}{\sqrt{m(x)}}\bigg)+V(x)\bigg]\psi(x,t)=i \hbar \frac{\partial}{\partial t}\psi(x,t).
\end{eqnarray}
Then, we take the Dunkl operator in the form of
\begin{eqnarray}\label{2}
D_{x_{i}}^{\mu_{i}}=\frac{\partial}{\partial x_{i}}+\frac{\mu_{i}}{x_{i}}(1-R_{i}),
\end{eqnarray}
where $R_i$ is the reflection operator which satisfies   
\begin{eqnarray}\label{2-1}
R_{i}f(x_{i})=f(-x_{i}).
\end{eqnarray}
Here, $\mu_{i}>-\frac{1}{2}$ are real numbers. We consider the following PDM \cite{35} 
\begin{eqnarray}\label{3}
m(x)=\frac{a^{2} m_{0}}{a^{2}+x^{2}},
\end{eqnarray}
 where $a>0$ is some parameter with the dimension of length. Next, we perform the gauge transformation, $\psi(x,t)=e^{-iEt}\psi(x)$, which leads to 
\begin{eqnarray}\label{4}
&&\frac{\left(a^{2}+x^{2}\right)}{a^{2}m_{0}}\frac{d^{2}\psi}{dx^{2}}+\left(\frac{2}{a^{2}m_{0}}x+\frac{1}{a^{2}m_{0}x}+\frac{2 \mu}{a^{2}m_{0}}\left(\frac{a^{2}+x^{2}}{x}\right)\right)\frac{d\psi}{dx}+\nonumber\\&&
\left(\frac{1}{a^{2}m_{0}}-\frac{\mu(1-R)}{m_{0}x^{2}}+\frac{2 \mu}{a^{2} m_{0}}\right)\psi=0
\end{eqnarray}
with change $x^{2}=-a^{2}z$ and considering $V(x)=0$, we have
\begin{eqnarray}
\frac{d^{2}\psi}{dz^{2}}+\bigg[\frac{\alpha_{1}+\alpha_{3}}{z}+\frac{\alpha_{1}+\alpha_{2}}{1-z}\bigg]\frac{d\psi}{dz}+\bigg[\frac{\alpha_{4}+\alpha_{5}}{z}+\frac{\alpha_{4}+\alpha_{5}}{1-z}+\frac{\alpha_{5}}{z^{2}}\bigg]\psi=0
\end{eqnarray}
where
\begin{eqnarray}
&&\alpha_{1}=-\frac{1}{2}\left(m_{0}+\frac{1}{a^{2}}\right),\quad \alpha_{2}=1,\quad \alpha_{3}=-\mu,\quad \alpha_{4}=\frac{1}{4}\left(1+2\mu\right)+\frac{E m_{0}a^{2}}{2 \hbar},\nonumber\\&& \alpha_{5}=\frac{\mu}{4}\left(1-R\right).
\end{eqnarray}

Next, we take Eq. \eqref{4} and express it in the following form

\begin{eqnarray}\label{3.15}
\tilde{H}=(z^{2}-z^{3})\frac{d^{2}}{dz^{2}}+\bigg[\left(\alpha_{2}-\alpha_{3}\right)z^{2}+\left(\alpha_{1}+\alpha_{3}\right)z\bigg]\frac{d}{dz}+\left(\alpha_{4} z+\alpha_{5}\right)
\end{eqnarray}

Employing Eq. \eqref{2.4}, we write the QES operator with the sl(2) generators as follows
\begin{eqnarray}\label{3.16}
	&&\tilde H = {C_{ +  + }}J_n^ + J_n^ +  + {C_{ + 0}}J_n^ + J_n^0 + {C_{ +  - }}J_n^ + J_n^ -  + {C_{0 - }}J_n^0J_n^ -  + {C_{ -  - }}J_n^ - J_n^ -  + {C_ + }J_n^ + 
	\nonumber\\&&+ {C_0}J_n^0 + {C_ - }J_n^ -  + C 
\quad\quad\end{eqnarray}
Substitution of Eq. \eqref{2.3} into Eq. \eqref{3.16} gives
\begin{equation}\label{3.17}
\left({P_4}\frac{{{d^2}}}{{d{z^2}}} + {P_3}\frac{d}{{dz}} + {P_2}\right)\tilde \Psi (z) = 0
\end{equation}
where 
\begin{subequations}\label{3.18}
\begin{eqnarray}
{P_4} &=& {C_{ +  + }}{z^4} +{C_{ +  0 }}{z^3}+ {C_{ +  - }}{z^2} + {C_{0 - }}z + {C_{ -  - }},\\
{P_3} &=& {C_{ +  + }}(2 - 2n){z^3} + ({C_ + } + {C_{ + 0}}(1 - \frac{{3n}}{2})){z^2}+ ({C_0} - n{C_{ +  - }})z \\&+& ({C_ - } - \frac{n}{2}{C_{0 - }}),\nonumber\\
{P_2} &=& {C_{ +  + }}n(n - 1){z^2} + (\frac{{{n^2}}}{2}{C_{ + 0}} - n{C_ + })z + (C - \frac{n}{2}{C_0}).
\end{eqnarray}
\end{subequations}
Comparing Eqs. \eqref{3.17} and \eqref{3.18} with Eq. \eqref{3.15}, we find

\begin{eqnarray}\label{3.23}
&&C_{++}=C_{0-}=C_{--}=C_{-}=0,\quad\quad C_{+0}=-1,\quad\quad C_{+-}=1, \quad\quad \nonumber\\
&& C_{0}=n+\alpha_{1}+\alpha_{3},\quad\quad C=\alpha_{5}+\frac{n}{2}(n+\alpha_{1}+\alpha_{3}),\\
&&C_{+}=\alpha_{2}-\alpha_{3}+1-\frac{3n}{2}\quad\quad -\frac{n^{2}}{2}-n C_{+}=\alpha_{4}. \nonumber
\end{eqnarray}

After performing the straightforward algebra, we obtain the energy eigenvalue function of the Dunkl-PDM Schrodinger equation in the form of 

\begin{eqnarray}\label{3.25}
E_{n}=\frac{2 \hbar}{m_{0}a^{2}}\left(n^{2}-n\left(2+\mu\right)-\frac{1}{4}\left(1+2\mu\right)\right)
\end{eqnarray}

Based on Eqs. \eqref{3.16} and \eqref{3.18}, we get the Lie algebraic differential operator, $\tilde{H}\in U_{sl(2)}$, as follows:

\begin{eqnarray}\label{3.24}
&&\tilde{H}=-J^{+}J^{0}+ J^{+}J^{-}+\left(\alpha_{2}-\alpha_{3}+1-\frac{3n}{2}\right) J^{+}+\left(n+\alpha_{1}+\alpha_{3}\right)J^{0}+\alpha_{5}\nonumber\\&&+\frac{n}{2}\left(n+\alpha_{1}+\alpha_{3}\right).
\end{eqnarray}

So, we have found the position dependent mass QES Dunkl-Schrodinger equation, this subspace retains the finite dimension $\tilde{P}_{n+1}$.  In the context of quasi-exactly-solvable, these limitations can be obtained with the assistance of the show theory. To this end, we construct the matrix eigenvalue problem of Eq.  \eqref{3.24} by considering 
\begin{eqnarray}\label{35}
\tilde{\psi}_{n}({\color{black}x})=\sum_{k=0}^{n}b_{k}{\color{black}x}^{k}\qquad,\qquad n=1,2,3,...
\end{eqnarray}
Then, by using Eq. \eqref{35} and Eq. \eqref{3.24}  with the sl(2) algebra, we  determine the exact solutions of the following cases, $n=0, 1$. After that,  we  generalize the solution to any $n$ case. 

\subsection{Solution for n=0}
In this case according to Eq. \eqref{35}, we take  $\tilde{\psi}_{0}(x)=\left\lbrace b_{0}\right\rbrace $. We observe a nontrivial solution if the potential parameters satisfy the following relation

\begin{eqnarray}
\alpha_{5}b_{0}=0,
\end{eqnarray}

that leads to the ground state energy eigenvalue to be found in the form of

\begin{eqnarray}\label{3.25a}
E_{0}=\frac{2 \hbar}{m_{0}a^{2}}\left(-\frac{1}{4}\left(1+2\mu\right)\right)
\end{eqnarray}

\subsection{Solution for n=1}
In the next case, we study $n = 1$. According to Eq. \eqref{35}, we take $\tilde{\psi}_{1}(x)=\left\lbrace b_{0}+b_{1}x\right\rbrace $, then the
corresponding matrix equation reads

\begin{align}\label{36}
{\left( \begin{matrix}
\alpha_{5}
	&0
	\\
	\alpha_{3}-\alpha_{2}
	& \alpha_{1}+\alpha_{3}+\alpha_{5}
	\end{matrix}\right) 
	\left( \begin{matrix}
	b_{0} 
	\\
	b_{1}
	\end{matrix}\right) }=0.
\end{align}

A nontrivial solution arises
\begin{eqnarray}\label{37}
\bigg(2\alpha_{5}(\alpha_{1}+\alpha_{3})+\alpha_{5}^{2} \bigg) b_{0}=0,
\end{eqnarray}

with the following relation

\begin{eqnarray}\label{38}
b_{1}=\left( \frac{\alpha_{2}-\alpha_{3}}{\alpha_{1}+\alpha_{3}+\alpha_{5}}\right) b_{0},
\end{eqnarray}

If we assume $b_{0}=1$, then  Eqs.  \eqref{38} and \eqref{35}  leads to the first excited-state wave-function

\begin{eqnarray}
\psi_{1}(x,t)=\frac{\exp\left({-iE t}\right)}{a^{2}}\left(-1-\left( \frac{\alpha_{2}-\alpha_{3}}{a^{2}\left(\alpha_{1}+\alpha_{3}\right)}\right)x^{2}\right)
\end{eqnarray}

After some algebra, we find that the wave-function for the n-th excited state can be expressed as follows:
\begingroup\makeatletter\def\f@size{11}\check@mathfonts
\begin{eqnarray}\label{p6}
\psi_{n}(x,t)=\exp({-iE t})\tilde{\psi}_{n}(x),
\end{eqnarray}
\endgroup
where $\tilde{\psi}_{n}(x)$ is
\begin{eqnarray}\label{55n}
\tilde{\psi}_{n}(x)=-\frac{1}{a^{2}}\sum_{k=0}^{n}b_{k}x^{2k}
\end{eqnarray} 
Here, the following recursion relation of coefficients has to be satisfied

\begin{eqnarray}
b_{k+1}=\bigg(\frac{k(k-1-\alpha_{2}+\alpha_{3}+\frac{k}{2})+(\alpha_{2}-\alpha_{3})-\frac{k^{2}}{2} }{(k+1)(k+\alpha_{1}+\alpha_{3})+\alpha_{5}}\bigg)b_{k},
\end{eqnarray}

with the boundary conditions $b_{-1}=0$. \color{black}

%\section*{Acknowledgements}
%{\color{red}The authors thank the anonymous reviewer for his/her helpful and constructive %comments. This work is supported by the Internal Project, [2022/...], of Excellent Research %of the Faculty of Science of University Hradec Kr\'alov\'e.}

\section{Conclusion}\label{sec4}
In this manuscript, we performed systematic work to obtain an exact solution of the PDM Dunkl-Schrodinger equation within the sl (2) algebra approach. The Dunkl operator that we introduced, has extra terms which take parity into account in addition to the normal derivative terms. This fact lets us investigate the problem within two states, even parity, and odd parity cases. In fact, the Dunkl operator is a modified form of the ordinary derivative and is a general form than its corresponding initial theory, and in the limited case of the Dunkl deformation parameter, the initial form of a normal derivative is recovered. Here, the important issue is the choice of the parameters $ \alpha $ and $ \mu$. It has been shown that the exact solvability of the PDM Dunkl-Schrodinger equation is dependent on these parameters and there is an interrelationship between the $\mu$ and $\alpha$ parameters. We, therefore, acquire constrained relations by using representational theory to solve this problem. Thus, by incorporating them, we were able to reach a general equation for energy, in which we compared our results in the special cases to those of previous works, and for $\alpha=\mu=0$ eigenvalues equal to rest energy. Adding Sl(2) algebra to the system, we found the wave functions of $n = 0$ and $1$ as well as arbitrary $n$.

\end{document}